\documentclass{jaa}

\usepackage{graphicx}

\def\vp{v_{\rm p}}
\def\vfi{v_{\phi}}
\def\bp{B_{\rm p}}
\def\bfi{B_{\phi}}
\def\qbr{Q_{\rm br}}
\def\qcyc{Q_{\rm cycl}}
\def\vpa{v_{\rm pA}}
\def\roa{\rho_{\rm A}}
\def\bpa{B_{\rm pA}}
\def\rd{r_{\rm d}}
\def\rod{\rho_{\rm d}}
\begin{document}

\title{Study of Magnetized accretion flow with cooling processes}


\author{Kuldeep Singh\textsuperscript{1,2} and Indranil Chattopadhyay\textsuperscript{1}}
\affilOne{\textsuperscript{1}Aryabhatta Research Institute of Observational Sciences 
(ARIES), Manora Peak, Nainital-263002, India.\\}
\affilTwo{\textsuperscript{2}University of Delhi, Delhi, India.}


\twocolumn[{

\maketitle

\corres{indra@aries.res.in}

\msinfo{1 January 2015}{1 January 2015}{1 January 2015}

\begin{abstract}
We have studied shock in magnetized accretion flow/funnel flow in case of neutron star with
bremsstrahlung cooling and cyclotron cooling. All accretion solutions terminate with a shock
close to the neutron star surface, but at some region of the parameter space, it also
harbours a second shock away from the star surface. We have found that cyclotron cooling is 
necessary for correct accretion solutions which match the surface boundary conditions.
\end{abstract}


}]


\doinum{12.3456/s78910-011-012-3}
\artcitid{\#\#\#\#}
\volnum{123}
\year{2016}
\pgrange{23--25}
\setcounter{page}{23}
\lp{25}

\section{Introduction}
\label{sec:intro}
Neutron stars (NS) are very fascinating objects because they have strong magnetic and gravitation fields.
Neutron stars accrete rotating matter from the companion, in the form of a disc, up to the distance where
$B^{2}/8\pi \sim \left(p,~\rho v^{2}\right)$. Within this radius, the magnetic field is so strong that it
channels all the matter along it and accretes in the form of
accretion curtains. These accretion curtains fall on the polar caps of the NS which is also known as funnel
flow/magnetized accretion. There are many existing models (e.g. Pringle and Rees 1972; Lamb {\em et al.} 1973; Davidson
and Ostriker 1973; Ghosh and Lamb 1979; Lovelace {\em et al.} 1985, 1986; Camenzind 1990; Paatz and Camenzind 1996; Ostriker
and Shu 1995; Li {\em et al.} 1999 etc) which address some of the issues of magnetized accretion flow, for e.g., inner radius of
accretion disc, transition zone and radiations from polar caps. However, most of the models are qualitative descriptions of the magnetized
accretion flow and related phenomena.  Koldoba {\em et al.} (2002) solved the equations of motion of matter falling
along the funnel in strong field approximation and in Newtonian gravity. But accretion solutions from their
model did not satisfy the surface boundary conditions, i.e., velocity near the star surface $\sim 0.1c$. Karino {\em et. al.} (2008)
studied possibility of shock formation in magnetized
accretion flow onto NS following Koldoba {\em et. al} (2002), but they considered only those solutions which satisfied the inner boundary
condition on the NS surface. In Karino {\em et. al.} (2008), accretion solution only satisfy
boundary conditions when shock location is very far from the NS surface, which is not generally the case.
Also in these models, they used Newtonian
gravity without any cooling mechanism. We know that Newtonian gravity fails when it is very close to compact objects like NS and cooling should be important in accretion column of a NS.
In this paper, we have studied magnetized accretion flow/funnel flow, by extending Koldoba {\em et. al.} (2002) and Karino {\em et. al.} (2008)
by considering Paczy\'nski-Wiita pseudo-potential to mimic strong gravity and cooling processes (bremsstrahlung and cyclotron cooling).

In section 2, we present the general magneto hydrodynamic (MHD) equations, assumptions, cooling processes and shock conditions.
The methodology to find the accretion
solution is explained in section 3. The parameter space and accretion flow solutions for rotation period $1$ s and $10$ ms are discussed in
section 4 and we give the concluding remarks in section 5. 
\section{Governing equations and assumptions}
\subsection{Basic Equations}
\label{sec:basiceqn}
In this study we have used ideal MHD equations. There are five conserved quantities which can be obtained
from MHD equations by integrating them along the field lines with steady state and axis-symmetry assumption.
These field lines are labelled by stream function of the magnetic field and it remains constant on a field line (Ustyugova {\em et al.} 1999,
Heinemann \& Olbert 1978). The MHD equations for steady state are
\begin{equation}
\nabla\ldotp(\rho \textbf{v}) = 0,
\label{conti}
\end{equation}
\begin{equation}
\nabla\ldotp\textbf{B} = 0,
\label{delB}
\end{equation}
\begin{equation}
\nabla\times(\textbf{v}\times \textbf{B}) = 0,
\label{fara}
\end{equation}
\begin{equation}
(\rho\textbf{v}\ldotp\nabla)\textbf{v} = - \nabla p + \frac{1}{c}(\textbf{J}\times\textbf{B}) + \Phi^{\prime}(r)\textbf{\^{r}}.
\label{mom}
\end{equation}
Here, $\rho$ is the mass density, ${\bf v}$ is the flow velocity, ${\bf B}$ is the magnetic field, $p$ is the fluid pressure,
$\Phi$ is the gravitational potential, $r$ is the radial coordinate and ${\hat {\bf r}}$ is the unit vector along $r$.
The velocity and magnetic field have poloidal and azimuthal components and are given by
${\bf v}\equiv(\vp,0,\vfi)$ and ${\bf B}\equiv (\bp,0,\bfi)$.
The angular velocity of matter is related to its azimuthal velocity as $\omega=v_{\phi}/r$.
In addition, the first law of thermodynamics or the conservation of energy equation is given by
\begin{equation}
 \rho \vp \left[\frac{d(e/\rho)}{dr}-\frac{p}{\rho^2}\frac{d\rho}{dx} \right]=\qbr+\qcyc,
\label{therm1law.eq}
 \end{equation}
where, $e=p/(\gamma -1)$ is the internal energy density and $\qbr$ and $\qcyc$ are the bremsstrahlung and cyclotron cooling terms.
The cooling terms are given by the brem-sstrahlung cooling
\begin{equation}
\qbr=\Lambda_{\rm br}\rho^{2}T_e^{1/2},
\label{colbr.eq}
\end{equation}
and cyclotron cooling is given by,
\begin{eqnarray} 
\nonumber
\qcyc=\Lambda_{\rm cycl}\left(\frac{A_{p}}{10^{15}\mbox{ cm}}\right)^{-17/40}\left(\frac{B_{p}}{10^{7}\mbox{ G}}\right)^{57/20}~~~~~~~~~~~~~\\
\times\left(\frac{\rho}{4\times 10^{-8}\mbox{ g}/\mbox{ cm}^{3}}\right)^{3/20}\left(\frac{T_e}{10^{8}\mbox{ K}}\right).~~~~~~~~~~~~~~~
\label{colcycl.eq} 
\end{eqnarray}
In the above, $\Lambda_{\rm br}\sim 5\times 10^{20}\mbox{erg}\mbox{ cm}^{-3}\mbox{g}^{-2}\mbox{K}^{-1/2}\mbox{s}^{-1}$ and
$\Lambda_{\rm cycl}\sim1.2\times 10^{8}\mbox{ erg}\mbox{ cm}^{-3}\mbox{s}^{-1}$ (Busschaert {\em et al.} 2015).
$T_e$ is the electron temperature and is, in general, smaller than the proton temperature, and we approximate
it as $T_e=\sqrt{m_e/m_p}T$ (Chattopadhyay \& Chakrabarti 2002, Das \& Chattopadhyay 2008).

The differential equations (\ref{conti})-(\ref{therm1law.eq}), when integrated, admit constants of motion and are given by
\begin{enumerate}
\item[(i)] From the continuity equation (\ref{conti}), we obtain
\begin{equation}
\rho \vp A_{\rm p} = {\rm constant} = {\dot M}=\mbox{ mass inflow rate},
\label{conMp.eq}
\end{equation}
where $A_{\rm p}$ is area of cross-section of flux tube.
\item[(ii)] Equation (\ref{delB}) gives the magnetic flux conservation,
\begin{equation}
 \bp A_{\rm p} = {\rm constant},
 \label{conBFp.eq}
\end{equation} 
and if we combine equations (\ref{conMp.eq}) and (\ref{conBFp.eq}), we can write $v_{p}$ as a function of $\rho$ and $B_{p}$,
\begin{equation}
 \vp=\frac{\kappa(\Psi)}{4\pi\rho}\bp,
 \label{vpBp.eq} 
\end{equation} where $\kappa$ is ratio of the mass flux to the magnetic flux and $\Psi$ is the stream function.
\item[(iii)] From Faraday equation (\ref{fara}),
\begin{equation}
 \Omega\left(\Psi\right) = \omega - \frac{\kappa(\Psi) \bfi}{4\pi\rho \varpi} = {\rm constant}.
 \label{conTBp.eq}
\end{equation}
Here, $\Omega$ is the angular velocity of the magnetic field, $\varpi =r {\rm sin}\theta$, where $\theta$ is the polar angle.
\item[(iv)] The azimuthal component of momentum balance equation (\ref{mom}) gives the total angular momentum $(\Lambda)$ conservation
which is the sum of the angular momentum of matter and angular momentum associated with the magnetic field of the star,
\begin{equation}
\Lambda(\Psi) = \omega \varpi^{2} - \frac{B_{\phi}\varpi}{\kappa(\Psi)} = {\rm constant}.
\label{conAngp.eq}
\end{equation}
\item[(v)] By integrating radial component of momentum balance equation (\ref{mom}), with the help other
equations, we obtain conservation of total energy $(E)$ along the field lines,
\begin{equation}
\begin{split}
E(\Psi) = \frac{1}{2}\vp^2 + \frac{1}{2}(\omega-\Omega)^{2}\varpi^{2} + h + \Phi(r) - \frac{\Omega^{2}\varpi^{2}}{2}  \\
 -\int\frac{Qdr}{\rho \vp} = {\rm constant}.
 \label{conEngp.eq}
 \end{split}
\end{equation}
where, $\Phi(r)=-\frac{GM}{r-r_{\rm g}}$ is Paczy\'{n}ski-Wiita potential $r_{\rm g}=\frac{2GM}{c^{2}}$ and $Q$ is total\
cooling which is given by
\begin{equation}
\nonumber
Q=\qbr+\qcyc.
\label{totcul.eq}
\end{equation} 
\end{enumerate}

\subsection{Assumptions}
\label{sec:assump}
We know that NSs have very strong magnetic field, so we can assume that
$\bp^2/8\pi\gg\left(p,~\rho v^{2}\right)$ (see Koldoba {\em et al.} 2002).
It means that dynamics of accretion process is controlled
by the magnetic field, which also implies that the flow is sub-Alfv\'enic, i.e. $\vp/\vpa \ll 1$
or $\roa/\rho\ll 1$, where $\vpa=\bpa/(4\pi\roa)$ is Alfv\'en speed and $\roa$
is mass density at the Alfv\'en radius. We assume that star has dipole like-magnetic field whose magnetic moment $(\mu)$ is aligned with the rotation axis of star, which can be
described by a stream function or flux function 
\begin{equation}
\Psi(r)=\frac{\mu}{r}\mbox{sin}^{2}\theta,~~\mbox{or,}~~r=\rd(\Psi)\mbox{sin}^{2}\theta.
\end{equation}
$\rd$ is the radius from where the matter starts channelling along the magnetic field lines.
By using these assumptions, from equations (\ref{vpBp.eq})-(\ref{conAngp.eq}), we can obtain two relations
(for more details, see Koldoba {\em et al.} 2002)
\begin{equation}
\frac{|\omega - \Omega|}{\Omega}\ll 1~~\mbox{and}~~\frac{\bfi}{\bp}\ll 1.
\label{omgB.eq}
\end{equation}
The first relation shows that angular velocity of matter is equal to the angular velocity of the field lines, i.e., matter and field lines are
co-rotating. Addition to this, if there is no slippage between the lines and the star's surface, i.e., field lines are frozen into the star's
surface, then we can equate $\Omega=\Omega_{star}$. The second relation implies that the azimuthal component of magnetic field $\bfi$ is
negligible as compared to the poloidal component of magnetic field $\bp$, and hence we can neglect $\bfi$ in our calculation. 

\subsection{Bernoulli function} 
\label{sec:bernfunc}
The Bernoulli function (equation \ref{conEngp.eq}) reduces to a simple form with the help of above mentioned assumptions and the
two relations in (\ref{omgB.eq}). Hence 
\begin{equation}
{\cal B}\left(r,\rho\right)=\frac{\vp^2}{2}+h+\Phi_{g}(r)-\int\frac{Qdr}{\rho \vp},
\label{bernl.eq}
\end{equation}
where $\vp=[{\mu\kappa(\Psi)}\left(4-3r/\rd\right)^{1/2}]/({4\pi\rho r^{3}})$, 
$h$ is enthalpy, $r_{co}=\left({GM}/{\Omega^{2}}\right)^{1/3}$ is the co-rotation radius, $\alpha=r_{co}/\rd$ is considered to be one in this
paper and 
$$
\Phi_{g}(r)=
-{\Omega}^{2}r^{2}_{co}\left(\frac{\alpha \rd}{r-r_{g}} + \frac{r^{3}}{2{\alpha}^{2}r^{3}_{d}}\right).
$$ 


The Equation of motion (EoM) is given by
\begin{eqnarray}
\frac{\partial {\cal B}}{\partial r}+\frac{\partial {\cal B}}{\partial \rho}\frac{d\rho}{dr}=0
~~\mbox{or}~~\frac{d\rho}{dr}=\frac{-{\partial {\cal B}}/{\partial r}}{{\partial {\cal B}}/{\partial \rho}}.   
\label{eom1.eq}
\end{eqnarray}
The flow starts with subsonic velocity at the $\rd$, but as matter moves towards the star, at some radius (say $r_{c}$), the
flow becomes super-sonic. At that radius $r_c$, the numerator and denominator of EoM (\ref{eom1.eq}) become zero or
$\frac{\partial {\rho}}{\partial r}|_{r_{c}}=\frac{0}{0}$. That is why $r_{c}$ is also known as the critical radius or the critical point.
However, $\frac{\partial {\rho}}{\partial r}|_{r_{c}}=\frac{0}{0}$ can be solved with L'Hospital's rule and we can obtain the
solution by integrating the EoM from the critical point. 

If there is no source or sink in energy equation, i.e., $Q=0$ in equation (\ref{therm1law.eq}), then we obtain the adiabatic relation
by integrating it and obtain 
\begin{equation} 
p={\cal K}\rho^{\gamma},
\label{vphip}
\end{equation}
where $\gamma ~( = 5/3)$ and $\cal K$ are adiabatic index and measure of entropy, respectively.  
If we use the continuity equation (\ref{conMp.eq}) and replace $\rho$ in terms of the sound speed,
then we can write the entropy-accretion rate as ${\dot {\cal M}}$ for adiabatic flow, 
\begin{equation} 
\dot{\cal M}=\vp A_{\rm p}c^{2n}_{s}~~\mbox{and}~~n=\frac{1}{\gamma-1}.
\end{equation}
Here, $c_s=\sqrt{\gamma p/\rho}$ is the sound speed.
Since in this paper we are considering the cooling processes, therefore ${\dot {\cal M}}$ is not constant along the flow.

\begin{figure}[!t]
\includegraphics[width=1.0\columnwidth]{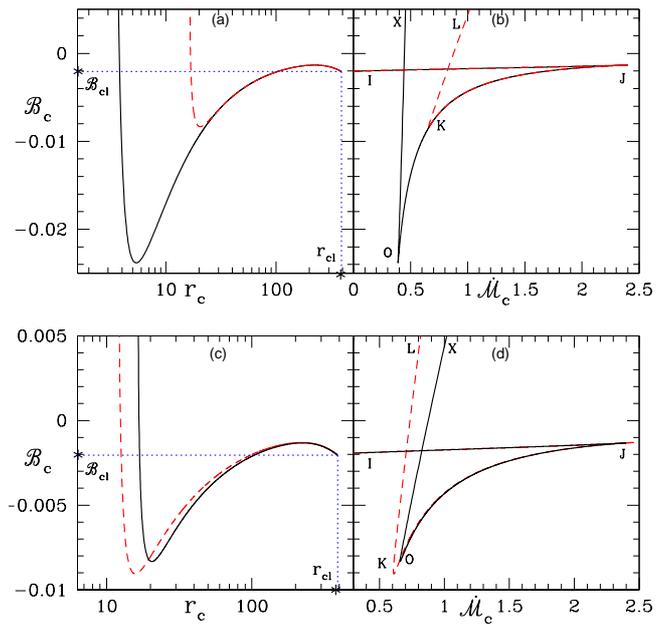}
\caption{\small Total energy ${\cal B}_c$ is plotted with $r_{c}$ \textbf{(a)} and \textbf{(c)}, and ${\cal B}_c$ versus entropy accretion rate
$\dot{\cal M}_{c}$ \textbf{(b)} and \textbf{(d)}. The plots are for two values of surface magnetic field
$B_{{\rm p}\circ}=10^{9}\mbox{ G}$ (solid-black line) and $10^{11}\mbox{ G}$ (dashed-red line) for a density
$\rod=5.0\times10^{-10}\mbox{ g}\mbox{ cm}^{-3}$ \textbf{(a)} and \textbf{(b)}. In lower two figures \textbf{(c)} and \textbf{(d)}, the plots are for two values of
density $\rod=5.0\times10^{-10}\mbox{ g}\mbox{ cm}^{-3}$ (solid-black) and $1.0\times10^{-8}\mbox{ g}\mbox{ cm}^{-3}$
(dashed-red line) for $B_{{\rm p}\circ}=10^{11}\mbox{G}$. All the plots are for $P=1$ s.}
\label{fig1}
\end{figure}

\subsection{Shock Conditions}
Shock conditions can be obtained from the conservation of fluxes, i.e mass flux conservation, momentum flux conservation
and total energy flux conservation (Kennel {\em et al.} 1989). If we use the strong magnetic field assumption and the two
relations in equation (\ref{omgB.eq}), then MHD shock conditions reduce to hydrodynamics shock conditions but the information of
magnetic field comes through the poloidal velocity and mass density. Therefore the shock conditions are
\begin{eqnarray}
\nonumber
\left[\rho \vp \right]=0,\\
\left[\rho \vp^2 + p\right]=0,\\
\nonumber
\left[\rho \vp \frac{\vp^2}{2} + \rho \vp h - \rho \vp \int\frac{Qdr}{\rho v_{p}}\right]=0.
\label{shk.eq}
\end{eqnarray}

\section{Methodology}
To find an accretion solution, we need five input parameters. Out of five parameters there are two main parameters,
rotation period $P$ of the star and density $\rod$ at $\rd$ radius. The third parameter is the star's surface
magnetic field $B_{{\rm p}\circ}$ which is related to rotation period of the star (Pan {\em et al.} 2013). The other
two parameters are the mass and radius of the star. By using these input parameters in the critical point conditions
which requires the numerator and denominator of EoM to be zero at the critical point, one can obtain the mathematical boundary condition
of the equations of motion. The EoM (equation \ref{eom1.eq}) can be recasted as
\begin{equation}
\frac{d v_{p}}{d r}=\frac{N(r,v_{p})}{D(r,v_{p})},
\label{eom2.eq}
\end{equation}
where, at $r=r_c$, or at the sonic point,
\begin{equation}
\nonumber
 N=\frac{3c^{2}_{sc}}{2r_c}\left(\frac{8-5r_c/r_d}{4-3r_c/r_d}\right)-\frac{\delta}{n}-\Phi^{'}_{g}(r_c)=0,
\end{equation}
and
\begin{equation}
\nonumber
D=\left(v_{{\rm p}c}-\frac{c^{2}_{sc}}{v_{{\rm p}c}}\right)=0.
\end{equation}
In the above,
\begin{equation}
\nonumber
\delta\equiv \frac{Q}{\rho v_{{\rm p}c}}.
\end{equation}
If we use the relation, $\vp={\mu\kappa(\Psi)}\left(4-3r/\rd\right)^{1/2}/({4\pi\rho r^{3}})$
 (because, $v_{p}={\kappa(\Psi)}\bp/[{4\pi\rho}]$) then we can rewrite $\delta$ as
\begin{equation}
\nonumber
\delta\equiv \frac{4\pi Qr^{3}_c}{\mu\kappa\sqrt{4-3r_c/r_d}}
\end{equation}

By solving these equations, we can determine $\rho_c$ (or $v_{{\rm p}c}$) and $T_c$ at $r_c$.
The total energy can be obtained by plugging these variables in equation (\ref{bernl.eq}) and
expressing ${\cal B}_c$, or ${\cal B}$ as a function of $r_c$.
The density slope or velocity slope at the critical point is calculated by using the L'Hospital's rule.
In the strong gravity domain, the centrifugal force (due to non-zero angular momentum of matter)
can modify the gravitational interaction and give rise to three types of critical points:
spiral type (or middle critical point), inner critical point and outer critical point, which are X-type
critical points. Suppose we have one X-type critical point
(either inner or outer critical point) and we have values of all the variables at the critical point. Then we
integrate the solution forward and backward from the critical point with fourth order Runga-Kutta method.
We simultaneously check the shock conditions (\ref{shk.eq}) in the supersonic branch.
If they satisfy the shock conditions at some radius, we obtain shocked accretion solution.
We calculate the post shock variables from these conditions and then start the integration
from there. So the next step is to iterate the shock location to find a solution which satisfies the surface
boundary conditions of the star. By using these methods, we have studied the parameter space and we have found many
consistent accretion solutions for different rotation periods of the star.      
\begin{figure*}[!t]
\centering
\includegraphics[width=1.6\columnwidth]{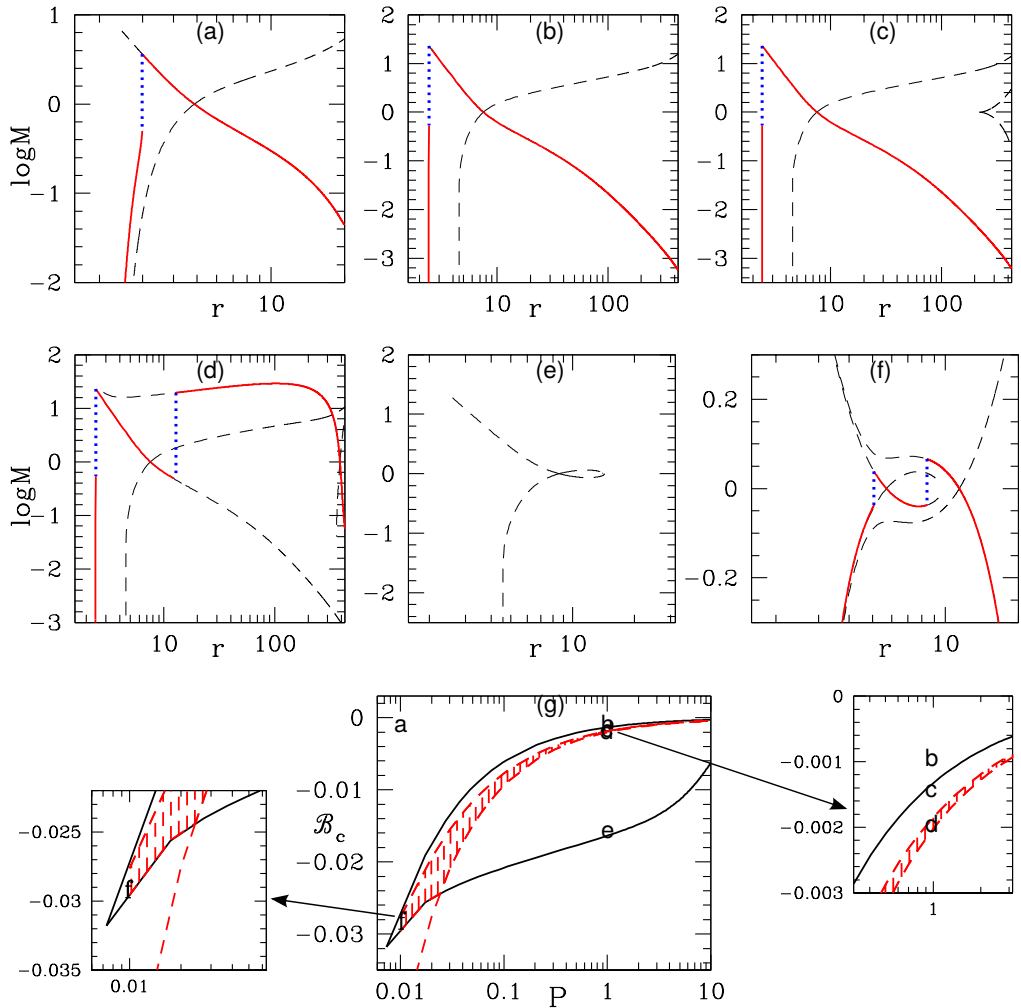}
\caption{\small Mach number $(M)$ versus $r$ for six solutions \textbf{(a)}-\textbf{(f)}. Each panel
corresponds to regions marked as \textbf{a}--\textbf{f} in ${\cal B}_c$--$P$ parameter space for MCP (solid bounded region in panel (g)).
In panels (a) to (f), accretion solution is represented by solid-red line, shock by dotted-blue line and other branches of
the solution by dashed-black line. In panel (g), region bounded by solid-black curve is the MCP region and shaded-red represent
the multiple shock region. All plots are for $\rod=5\times 10^{-10}$ g cm$^{-3}$.}
\label{fig2}
\end{figure*}

\section{Results and discussion}
An accretion disc with Keplerian angular momentum distribution, extending from large distance
to the star surface is a possibility, however, in presence of magnetic field this is a remote possibility.
Matter would channel along magnetic field lines onto the poles. And the surface of the star at the poles
will resist the supersonic flow. Therefore, an accretion on to a compact star with hard surface,
is necessarily associated with a shock transition close to the star surface. In this paper, we discuss
all possible accretion solutions in steady state, falling along the magnetic field lines under the assumptions made in section \ref{sec:assump}.
All these solutions should terminate in a shock, 
In this paper, the analysis is done in the geometrical units, which means that velocity is in units of $c$ and distance in
terms of $r_g={2GM}/{c^2}$.  We have considered that the radius and mass of the star as $R_{\rm star}=1.0\times 10^{6}$cm
and $M=1.4$ $M_{\bigodot}$, respectively. Therefore, in units of $r_g$, the radius of the star is $R_{\rm star}=2.418r_g$.
Although, the analysis is done in geometric units, we quote the rotation period $(P)$ in seconds, in order to provide a feeling
of the rotation period of the star for the reader. 
As we discussed in the methodology section, for a set of input parameters, basically the star rotation period $P$,
$\rod$ and $B_{{\rm p}\circ}$, we can obtain mass density $\rho_c$, temperature $T_c$ and
${\rm B}_c$ at the critical point. So, for $P=1\mbox{ s}$ and $\alpha=1$, we have plotted the ${\cal B}_c$ versus
$r_c$ in Figures \ref{fig1}(a), (c) and ${\cal B}_c$ versus $\dot{\cal M}_c$ in Figures \ref{fig1}(b), (d).
In upper two panels of Figures \ref{fig1}(a) and 1(b), we consider
$\rod=5.0\times10^{-10}\mbox{g}\mbox{ cm}^{-3}$ but two values the magnetic field $B_{{\rm p}\circ}=10^{9}\mbox{G}$ (solid-black line)
and $B_{{\rm p}\circ}=10^{11}\mbox{G}$ (dashed-red line). In Figures \ref{fig1}(c) and \ref{fig1}(d),
we consider the surface magnetic field as $B_{{\rm p}\circ}=10^{11}\mbox{G}$ but two values of
$\rod=1.0\times10^{-8}\mbox{g}\mbox{ cm}^{-3}$ (solid-black line) and $\rod=5.0\times10^{-10}\mbox{g}\mbox{ cm}^{-3}$
(dashed-red line). 
If there is a maximum and a minimum in the ${\cal B}_c-r_c$ curve for a given value of $P$, then there is a possibility
of forming multiple critical points or MCP. The dotted lines in Figures \ref{fig1}(a) and 1(c) show the upper limit of $r_c$, i.e., $r_{cl}$ and
its corresponding energy ${\cal B}_{cl}$. Let us call the minimum (or maximum)
in ${\cal B}_c$ as ${\cal B}_{c\rm min}$ (or ${\cal B}_{c\rm max}$), then for
${\cal B}_{c\rm max}>{\cal B}>{\cal B}_{c\rm min}$,
the flow will harbour multiple sonic/critical points. But if ${\cal B}_{c\rm min}\leq {\cal B}\leq
{\cal B}_{cl}$, then there can only be two sonic points, where the inner one is the X-type
and the other one as the spiral type.
The ${\cal B}_c$---${\dot {\cal M}_c}$ plots (Figures \ref{fig1}(b) and 1(d))
also show the same
effect. The kite-tail feature is symptomatic of MCP. Various combinations of ${\cal B}_c$ and ${\dot {\cal M}}_c$,
produce the sonic points, the LK (XO) branch of the dashed (solid) curve is the loci of inner sonic points, the
KJ (OJ) branch of the dashed (solid) is loci of middle sonic points and IJ is loci of outer sonic points (Figures \ref{fig1}(b), 1(d)).
As long as the entropy of the inner sonic point (${\dot {\cal M}}_c$) is greater than that of the outer sonic point, there is possibility
of a second shock formation away from the star surface, in addition to the shock forming close to the star surface.

As has been pointed out, the accretion onto a compact star with hard surface should end with a shock very close to the surface.
In this paper, we aim to study all possible accretion solutions in addition to this terminal shock near the star surface.
The sonic point properties of Figures \ref{fig1}(a)-(d) appear similar to the black hole systems (Kumar \& Chattopadhyay 2013, 2014, 2017,
Chattopadhyay \& Kumar 2016) although the inner boundary condition is quite different. However, unlike
hydrodynamic studies in black hole system,
in the strongly magnetized NS system, we find that $r_c \leq r_{cl}$.
Observations also tells us that there exist a relation between period of rotation and the surface magnetic field (Pan et. al. 2013).
Thus, henceforth we consider a simple empirical relation inspired from the observations of Pan et. al. (2013), and we assume
it to be $B_{{\rm p}\circ}=10^{0.75logP+10}$, where $B_{{\rm p}\circ}$ and $P$ is in Gauss and in secs, respectively. 
If we now change $P$, we will get a set of ${\cal B}_c$---$r_c$ plots, each with its own extrema and ${\cal B}_{cl}$. If all the maxima
and minima are joined in the ${\cal B}_c$---$P$ parameter space, we obtain a bounded region (solid curve) in Fig. \ref{fig2}(g). 
Any set of parameters in this bounded region will produce MCP
for the accretion solution. The shaded region will produce the outer, or the second shock. The lower bound of the
shaded region is the loci of ${\cal B}_{cl}$. Any point within the lower bound of the shaded region and the lower bound of MCP region
will harbour two sonic points, the inner one being X-type and the outer one being spiral type.
Any parameter set between the lower bound of the
shaded region and the upper bound of the MCP region will harbour three sonic points, the inner and outer being X-type and the centre as
spiral type.
We have marked locations on the ${\cal B}_c$---$P$ parameter space as `a'-`f', and then plotted the corresponding solutions in terms of the
Mach number $M=\vp/c_s$ as a function of $r$, in Figures \ref{fig2}(a)-(f), named after the location of the corresponding
coordinates in Fig. \ref{fig2}(g).
The accretion solutions are represented by solid-red line,
shock is denoted by dotted-blue line and other branches of solution are represented by dashed-black line.
Figure \ref{fig2}(a) corresponds to high energy (${\cal B}_c=0.001$) and low rotation period ($P=10$ ms), the accretion flow becomes transonic through the
inner sonic, goes very close to the star surface and undergoes a shock transition.
The next solution is
Fig. \ref{fig2}(b) plotted for accretion parameters ${\cal B}={\cal B}_c=0.001$ and $P=1$ s which lies outside the bounded region or
MCP region (solid-black line). This solution is similar to the previous one.
The third solution is Fig. \ref{fig2}(c), for parameters $P=1$ s and ${\cal B}={\cal B}_c=0.0015$ which just lies inside the MCP 
region (see inset to Fig. \ref{fig2}(g)). Therefore in Fig. \ref{fig2}(c), we have one solution which passes through the inner critical 
point and other is the cusp type solution through the outer critical point (dashed).
Therefore the flow becomes transonic after crossing the inner sonic point and suffers only one shock near the star surface.
Figure \ref{fig2}(d) is the solution for $P=1$ s and ${\cal B}_c=0.00203$ which lies 
in the multiple shock region (shaded-red region). Therefore, in this case, flow first passes through the outer critical point and 
becomes super-sonic. Then it forms a shock transition before the inner critical point, it passes through the inner critical point and 
form another shock near the star surface to satisfy the surface boundary conditions.     
The solution in Fig. \ref{fig2}(e) is situated below the lower bound of the dashed region i.e., ${\cal B}< {\cal B}_{cl}$ 
and therefore shows an inverted $\alpha$ type solution. Since this is not a global solution, so no accretion is possible. 
In this case, we have only two critical points: one is X-type and other is
spiral type but we do not have outer critical point.
In all the above cases, the X-type sonic points are those were the accretion and wind type solutions cross, while the solutions
skirts the spiral type sonic point.
The last solution is Fig. \ref{fig2}(f),
for rotation period $P=10$ ms which is same as that in Fig. \ref{fig2}(a), but it lies inside the
multiple shock region (shaded-red). In this case also, flow has double shock, similar to the solution Fig. (\ref{fig2}d).
However in this case, solution which passes through the inner critical point is alpha type.

\begin{figure}[!t]
\includegraphics[width=1.0\columnwidth]{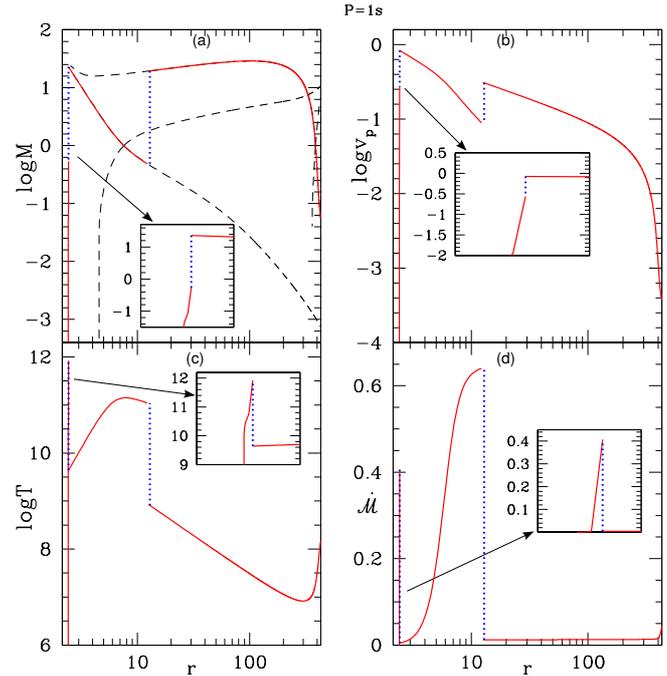}
\caption{\small The accretion solution for $P=1$ s and $B_{{\rm p}\circ}=10^{10}\mbox{ G}$. \textbf{(a)} $M$, \textbf{(b)} $\vp$, \textbf{(c)} $T$ and
\textbf{(d)} $\dot{\cal M}$ as a function of $r$. The accretion solution is represented by solid-red line, shock by dotted-blue line and
other branch of solution by dashed-black line. Accretion parameters same as in Fig. \ref{fig2}(d).}
\label{fig3}
\end{figure}

We have plotted the Mach number 
$M$ (Fig. \ref{fig3}(a)), poloidal velocity $\vp$ (Fig. \ref{fig3}(b)), temperature
$T$ (Fig. \ref{fig3}(c)) and entropy-accretion rate $(\dot{\cal M})$ (Fig. \ref{fig3}(d)) as a function of $r$.
The accretion parameters are $P=1$ s, $B_{{\rm p}\circ}=10^{10}\mbox{ G}$, $\rod=5\times 10^{-10}$ g cm$^{-3}$
and ${\cal B}=0.00203$ (similar to Fig. \ref{fig2}(d)). 
The accretion flow passes through the outer sonic point and then suffers a shock at $12.897~r_g$. The subsonic post shock matter
accelerates and again becomes supersonic as it crosses the inner sonic point. This supersonic matter again suffers
 a shock at $2.434~r_g$, which is very close to the surface of the star. The inset in each panel zooms on to the 
 inner shock. The temperature and the poloidal velocity starts from quite low values and becomes relativistic.
 The temperature in the inner post-shock disc reaches to a very high values, only decrease drastically down to very lower
 values. Because of the presence of cooling the entropy is not constant (Fig. \ref{fig3}(d)).
 
\begin{figure}[!t]
\includegraphics[width=0.8\columnwidth]{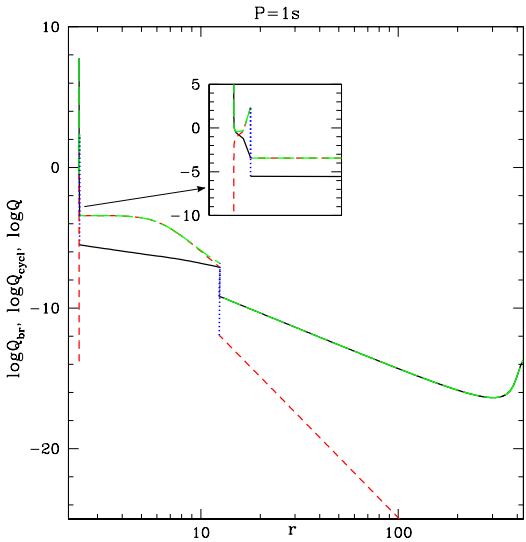}
\caption{\small Variation of emissivity of different radiative processes $\qbr$ (solid, black), $\qcyc$ (dashed, red),
and total cooling $Q$ (long-dashed, green) which corresponds to the solution as in Figures \ref{fig3}(a)-(d).}
\label{fig4}
\end{figure}

 In Fig. \ref{fig4}, we plot the emissivity of different cooling processes with $r$ for the accretion solution depicted
 in Figs. \ref{fig3}(a)-(d) for $P=1$ s. Different curves represent the bremsstrahlung $\qbr$ (solid, black),
 cyclotron $\qcyc$ (dashed, red) and the total emissivity $Q$ (long-dashed, green). At larger distance the
 cooling process is dominated by bremsstrahlung, while in the post-shock region and near the surface, cyclotron
 process dominate. The cooling rate close to the surface overwhelms and both $T$ and $\vp$ plummets several orders of magnitude.
 The inset zooms the cooling rates in the post-shock region close to the star surface.
 
\begin{figure}[!t]
\includegraphics[width=0.8\columnwidth]{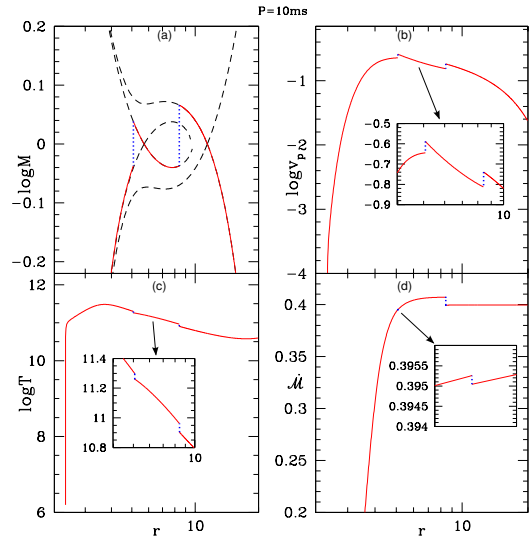}
\caption{\small Variation of \textbf{(a)} $log(M)$, \textbf{(b)} $log(\vp)$, \textbf{(c)} $log(T)$ and \textbf{(d)} ${\dot {\cal M}}$
with $r$. Accretion solution (solid, red), shock jump (dotted, blue) and other branches (dashed, black)
are shown. Other parameters are $P=10$ ms, $\rod=5\times 10^{-10}$ g cm$^{-3}$.}
\label{fig5}
\end{figure}

We plot $M$ (Fig. \ref{fig5}(a)), $\vp$ (Fig. \ref{fig5}(b)), $T$ (Fig. \ref{fig5}(c)) and ${\dot {\cal M}}$ (Fig. \ref{fig5}(d)),
which also harbours multiple shock with a much faster rotating system with $P=10$ ms,
$B_{{\rm p}\circ}=3.16\times10^{8}\mbox{ G}$ and $\rod=5\times 10^{-10}$ g cm$^{-3}$. This is the same solution as is depicted
in Fig. \ref{fig2}(f). 
The two shocks are located at $5.081$ and at $8.414$, but are much weaker than the previous case. 

\begin{figure}[!t]
\includegraphics[width=1.0\columnwidth]{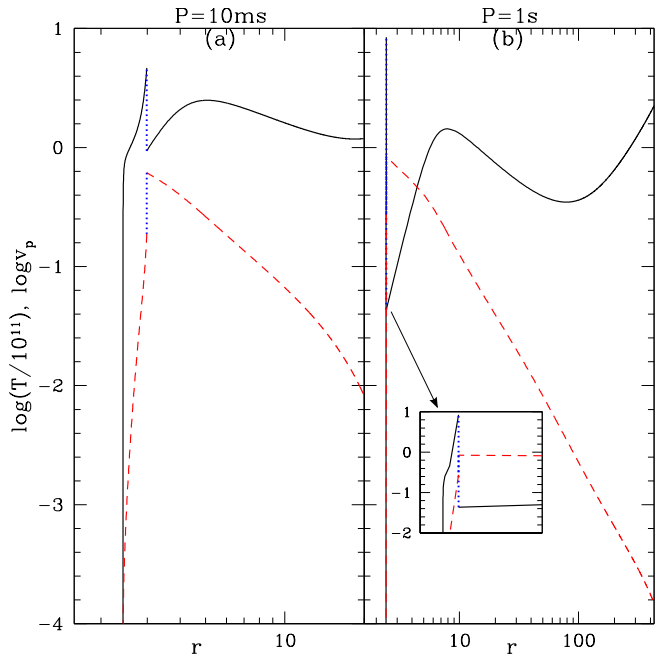}
\caption{\small Accretion solutions $log(\vp)$ (dashed, red) and $log(T)/10^{11}$ (solid, black) for $P=10$ ms \textbf{(a)} and $P=1$ s
\textbf{(b)}. Other accretion parameters are ${\cal B}=0.001$ and $\rod=5\times 10^{-10}$ g cm$^{-3}$.}
\label{fig6}
\end{figure}

In Figs. \ref{fig3}(a)-\ref{fig5}(d), we have discussed two solutions with high and low spin periods, which becomes
transonic at two places and harbour multiple shocks.
Now, we look into two solutions once again for high and low periods, but which becomes transonic only at one place,
and harbour one shock close to the star surface. 
As a representative case, we consider the two solutions depicted in Figures \ref{fig2}(a) and (b) characterized by spin
period $P=10$ ms and $P=1$ s, respectively, but with the same Bernoulli parameter ${\cal B}$. In Fig. \ref{fig6}(a),
we plot the poloidal velocity $\vp$ and temperature $T$ for $P=10$ ms in log scale. In Fig. \ref{fig6}(b),
we plot the poloidal velocity $\vp$ and temperature $T$ for $P=1$ s in log scale. In both the case,
the shock is at $3 r_g$, which is just outside the star surface, but the post-shock temperature is higher for the
slowly rotating star. The cyclotron radiation in the post-shock region is quite strong and drastically reduce both $\vp$ and $T$,
such that it satisfies the inner boundary condition of $\vp(R_{star})\sim 0$.

\section{Conclusion}
In this paper, we have studied the magnetized accretion solutions which have one or two shocks for different rotation period of star and including cooling mechanisms (bremsstrahlung and cyclotron) in the Paczy\'{n}ski \& Wiita potential. From our study of the MCP parameter space, for range of rotation period (few ms - 10s), we find that there is a lower limit on rotation period $P_{min}$ below which the MCP are not possible so we cannot have multiple shocks. Therefore, for accreting NS cases, where spin period is low e.g.,
SAX J1808.4-3658 with $P=2.41$ ms, there can be only one shock very close to the star surface. 

We find correct accretion solutions with cooling mechanisms (bremsstrahlung and cyclotron cooling) for different rotation period, $P=10\mbox{ ms}$ and $1\mbox{ s}$. 
Depending on the energy of the flow, accretion solutions may have one or two stationary shocks.
If two shocks are present then, the outer shock is at radius $\sim 10r_g$ and inner shock is at a radius $\sim 3r_g$,
else, only one shock is present very close to the surface. One has to consider cooling mechanism, otherwise
either the velocities are too high close to the star surface (Koldoba et. al. 2002) or the shocks
are much further away than what observations suggest.

Our results show that cooling processes dominate near the star. We found that bremsstrahlung emission dominates at larger distance, as well as near the star surface,
while cyclotron cooling dominates in a region closer to the star. In black hole system,
it has been noted that the bremsstrahlung process to be an inefficient cooling process, while in the
present NS case, we found this process to be quite effective. The simple
reason is that the, the cross-section of accretion on to a black hole is
spherical (i.e., $\propto r^2$, matter crosses horizon with $c$), while
for NS the cross section or $A_{\rm p}$ is dictated by the magnetic field configuration. 
Now, $dA_{\rm p}/dr$ for a dipole magnetic field, increases sharper than that
of spherical cross-section, which makes flow near NS surface to be denser.
This makes bremsstrahlung to be stronger
for NS case. As the result of the presence of cooling processes, all the solutions in this paper
satisfy $\vp \lesssim 10^{-8}c$ close the star surface. 
We must however accept that, since the rotation axis and magnetic axis are aligned in the present analysis therefore, it can
only make qualitative assessment of pulsars. This analysis is more suited to address accreting NS system which have weak pulsations
or none at all.

These accretion solutions with cooling mechanisms are able to connect the flow from the disc to the star surface via shock/shocks.
We find that cyclotron cooling is necessary along with bremsstrahlung cooling to find plausible accretion solutions. These solutions
should help to explain the accretion process and radiation mechanisms of accreting NSs.





\end{document}